\documentclass[runningheads]{llncs}

\usepackage{accv}

\usepackage{accvabbrv}      
\usepackage{graphicx}
\usepackage{booktabs}
\usepackage{amsmath}
\usepackage{amssymb}
\usepackage{multirow}
\usepackage{pgfplots}
\usepackage{subcaption}
\usepackage{float}
\pgfplotsset{compat=1.17}
\usepackage[accsupp]{axessibility}  

\usepackage{hyperref}
\usepackage{orcidlink}

\newcommand{\Lcnn}{L1$_{\text{CNN}}$}
\newcommand{\Llm}{L2$_{\text{LM}}$}
\newcommand{\Lvlm}{L3$_{\text{VLM}}$}

\begin{document}

\title{Rethinking Streaming-Perception Evaluation on Heterogeneous Edge Platforms}

\titlerunning{Rethinking Streaming-Perception Evaluation}

\author{Misun Yu\inst{1} \and Jinyoung Moon\inst{1} \and Jemin Lee\inst{2}}
\authorrunning{M. Yu et al.}
\institute{Electronics and Telecommunications Research Institute (ETRI), Daejeon, Republic of Korea\\
\email{\{msyu,jymoon\}@etri.re.kr}
\and
Division of Electronics and Information Engineering, Jeonbuk National University, Jeonju, Republic of Korea\\
\email{jemin.lee@jbnu.ac.kr}}

\maketitle

\begin{abstract}
Multi-camera streaming perception is increasingly deployed on heterogeneous edge platforms shared with co-resident workloads, yet accelerator placement is often evaluated using isolated single-stream experiments and mean streaming average precision (sAP). Using two end-to-end pipelines on a single GPU--NPU platform, we show that isolated evaluation can mis-rank deployment-time placement. Although the GPU pipeline is preferred in isolation, GPU-localized contention introduces deadline misses that make detections stale and can reverse the preferred placement before full GPU saturation. The NPU pipeline is less accurate than the GPU pipeline on small and medium objects in isolation, but nearly matches it on large objects. The largest absolute sAP losses in our latency and contention experiments occur for large objects. In our four-stream experiments, the preferred placement depends on which path becomes stale, and increasing GPU-side contention shifts the best placement from All-GPU to All-NPU. Under a GPU-saturating vision--language co-tenant, All-NPU achieves $5.2\times$ the worst-stream sAP of All-GPU. Because mean sAP can hide severe single-stream degradation, evaluation should report contention sweeps, deadline-miss rates on both paths, and worst-stream sAP alongside mean sAP.
\end{abstract}


\section{Introduction}

Autonomous vehicles and edge AI systems process multiple camera streams concurrently~\cite{argoverse,waymo}, and accuracy depends not only on detection quality but also on when detections become available. Streaming average precision (sAP) captures this timing effect by evaluating the latest available detections against the ground truth at each evaluation timestamp~\cite{streaming-perception}. Recent work has improved streaming detectors and latency--accuracy trade-offs~\cite{streamyolo,damo-streamnet,asap,longshortnet}, but evaluation is still commonly performed on a single detector running in isolation on a single accelerator and summarized with mean sAP.

Real deployments differ from this setting. Heterogeneous accelerators such as GPUs and NPUs share the platform with co-resident workloads~\cite{mobilint-aries,igniter}, but accelerator placement is often chosen based on isolated profiling, implicitly assuming that the same ranking holds under deployment-time contention~\cite{partitiontuner,ngs,rl-scheduler,tod}. We show that this assumption can fail. Comparing two end-to-end pipelines on our platform, we find that the FP32 GPU pipeline achieves higher sAP than the INT8 NPU pipeline in isolation. However, this ranking can reverse when contention is concentrated on the GPU path. Under such contention, longer processing times lead to deadline misses and stale detections on the GPU path. The decrease in sAP of the GPU pipeline can exceed that of the NPU pipeline by more than the sAP advantage of the GPU pipeline in isolation.

Mean sAP can further hide this failure, because contention may severely degrade one stream while aggregate accuracy appears acceptable. Rather than proposing a new detector or scheduler, we revisit the evaluation protocol for streaming perception as a basis for placement decisions under deployment-time contention. Our contribution is a deployment-faithful evaluation protocol for multi-camera streaming perception on shared heterogeneous platforms. We support this protocol with three empirical findings.

\begin{itemize}
\item \textbf{Isolated evaluation can mis-rank accelerator placement.}
The pipeline preferred in isolation can become inferior under deployment-time contention, and selection criteria based on isolated accuracy, isolated latency, or utilization thresholds miss this reversal.

\item \textbf{Accuracy differences in isolation and latency-associated sAP losses show contrasting size dependence.}
The accuracy advantage of the GPU pipeline in isolation is concentrated on small and medium objects, while the largest absolute sAP losses under latency and contention occur for large objects. The observed deployment-time ranking changes are consistent with differences in latency-associated losses between the two paths.

\item \textbf{Worst-stream sAP reveals failures hidden by mean sAP.}
Degradation can concentrate in a single camera stream, so worst-stream sAP is necessary alongside mean sAP.
\end{itemize}

\section{Related Work}

\paragraph{Streaming perception.}
Streaming perception evaluates the latest available detections against the ground truth at each time instant, making latency a first-class component of accuracy~\cite{streaming-perception}. Subsequent work has introduced streaming-oriented detector designs and evaluation protocols~\cite{streamyolo,damo-streamnet,asap,longshortnet}. Beyond aggregate sAP, an extended version of StreamYOLO~\cite{streamyolo-vsap} proposed velocity-aware sAP (VsAP), which averages sAP across simulated driving-speed settings obtained by changing frame-sampling intervals. This evaluates streaming accuracy across simulated driving speeds, whereas we analyze differences across object sizes. More recent methods treat delay as a design variable, predicting future frames~\cite{transtreaming} or compensating latency in 3D detection~\cite{lasp}. These studies typically evaluate a single detector running in isolation on one accelerator.

\paragraph{Heterogeneous edge inference.}
Modern edge platforms combine heterogeneous accelerators and co-resident workloads to balance throughput, latency, and power consumption~\cite{mobilint-aries}. Prior work selects device placements and schedules using isolated accuracy, isolated latency~\cite{tod,partitiontuner,ngs,rl-scheduler}, and measured or predicted GPU utilization, which is a common scheduling signal in GPU cluster management~\cite{gandiva,horus}. Interference-aware systems such as iGniter~\cite{igniter} further predict contention-induced latency, but they target latency SLOs rather than delivery-time streaming accuracy and do not consider heterogeneous GPU--NPU placement. We neither claim that such schedulers generally fail nor propose one. Streaming perception thus measures delivery-time accuracy but usually assumes a fixed accelerator, while heterogeneous edge inference optimizes placement but evaluates static accuracy or latency. We instead study deployment-time streaming accuracy under accelerator sharing and evaluate the placements implied by common selection criteria.

\paragraph{Worst-case evaluation.}
Robust learning and large-scale systems emphasize worst-case behavior, including worst-group accuracy and tail latency~\cite{groupdro,cvar,tailatscale}, and multi-stream video analytics has explored optimizing minimum accuracy across streams~\cite{biswift,videoedge}. Streaming-perception evaluation, however, typically reports aggregate sAP, which can hide severe degradation in a single camera stream. We therefore use worst-stream sAP alongside mean sAP.


\section{Problem Formulation}
\label{sec:problem}

Consider multiple camera streams running concurrently on a heterogeneous edge platform, each with a detector placed on one of two end-to-end pipelines: an FP32 GPU pipeline or an INT8 NPU pipeline. The goal is to maximize deployment-time streaming accuracy under the actual co-resident workload. We ask when the pipeline ranking observed in isolation can reverse after deployment, what mechanism causes the reversal, and whether mean sAP exposes the resulting degradation.

\begin{figure}[tb]
\centering
\includegraphics[width=\linewidth]{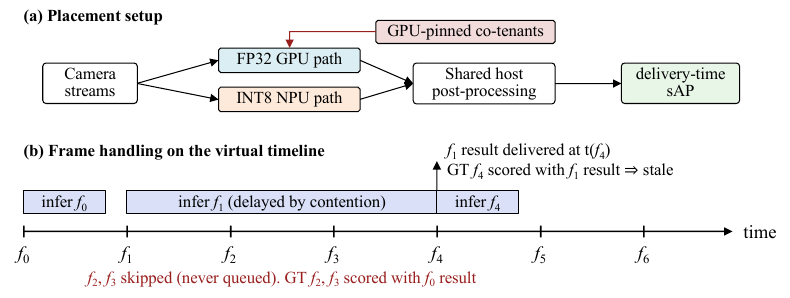}
\caption{Overview of the evaluated system. (a) GPU--NPU placement with GPU-pinned co-tenants and shared host post-processing. (b) Frame skipping and evaluation with stale detections when processing is delayed by contention.}
\label{fig:overview}
\end{figure}

\paragraph{Definitions.}
\emph{Contention} is competition for shared resources among co-resident workloads, which can lengthen foreground processing, including host-side stages. \emph{Staleness} is the temporal mismatch between the scene reflected by the latest delivered detection and the ground truth at evaluation time. Figure~\ref{fig:overview} illustrates the evaluated system and how processing delays produce stale detections under the frame-handling protocol. A \emph{reversal} occurs when the placement preferred without co-tenants is outperformed by another placement under contention, with the evaluation metric and stream configuration held fixed.

\paragraph{Modeling placement reversal.}
We describe deployment-time accuracy per pipeline $p\in\{\mathrm{GPU},\mathrm{NPU}\}$ and per object-size category $b \in \{\mathrm{small}, \mathrm{medium}, \mathrm{large}\}$, because both the difference in sAP between the two pipelines without co-tenants and the losses under contention can vary by object size. Let $a_p(b)$ denote the sAP of pipeline $p$ measured without co-tenants under the same evaluation configuration, and let $C$ denote the contention level. Under contention $C$,
\begin{equation}
\mathrm{sAP}_{p}(b,C) = a_p(b)-S_{p}(b,C),\qquad G(b) = a_{\mathrm{GPU}}(b)-a_{\mathrm{NPU}}(b),
\label{eq:npu}
\end{equation}
where $S_p(b,C)$ is the observed sAP loss under contention $C$ relative to the uncontended value $a_p(b)$ and may be negative. The difference $G(b)$ is the sAP of the GPU pipeline minus that of the NPU pipeline without co-tenants, referred to as the isolated pipeline gap. This difference does not separate device, precision, and software-stack effects. The NPU pipeline attains higher sAP for size category $b$ when
\begin{equation}
S_{\mathrm{GPU}}(b,C) - S_{\mathrm{NPU}}(b,C) > G(b).
\label{eq:crossover}
\end{equation}
Without co-tenants, both $S_p$ are zero, so the GPU pipeline has higher sAP when $G(b)>0$. A reversal occurs when the NPU pipeline attains higher sAP under contention despite this isolated GPU advantage. This relation characterizes the reversal condition for single-stream results in each object-size category but does not predict the contention level at which it is met.

\section{Deployment-Faithful Evaluation of Placement Decisions}

We evaluate whether isolated pipeline profiling identifies the best accelerator placement under co-resident workloads. We examine isolated pipeline accuracy, size-dependent sAP losses, placement ranking changes under contention, and degradation revealed by worst-stream sAP.

\subsection{Experimental Setup}
\label{sec:setup}

Each experiment varies the number of foreground camera streams, the GPU co-tenant load, and the GPU--NPU placement. CPU execution is excluded because it drops most frames under co-tenancy.

\paragraph{\textbf{Platform and runtimes.}}
All experiments run on a single host with an Intel Core Ultra 9 285K CPU (24 cores), an NVIDIA RTX~5090 GPU, and a Mobilint MLA100 NPU~\cite{mobilint-aries}. Foreground inference uses FP32 PyTorch on the GPU and vendor-compiled INT8 models on the NPU, which supports up to eight concurrent detector instances. Host-side post-processing (dequantization, decoding, NMS) uses four threads in all placement and contention experiments. In the default experiments with four foreground streams ($N{=}4$), the four foreground threads share the default CUDA stream. The QoS experiment (Sec.~\ref{sec:results:reversal}) instead uses one shared maximum-priority stream. In a separate experiment on the same NPU pipeline, we vary the number of host-side post-processing threads to examine the effect of processing latency on sAP. We refer to this experiment as the same-NPU latency control (Sec.~\ref{sec:results:single}).

\paragraph{\textbf{Detectors, dataset, and stream composition.}}
We use COCO-pretrained YOLO11s~\cite{yolov11} as the foreground detector in all main experiments. In the main experiments, both paths resize each $1200\times1920$ frame to $400\times640$ and then pad it to $416\times640$ (GPU) or $640\times640$ (NPU). To assess generality, we additionally evaluate YOLOv8n and YOLOv8s~\cite{yolov8}, YOLOv10s~\cite{yolov10}, and YOLO12s~\cite{yolov12}. All detectors are evaluated on the Argoverse-HD validation split~\cite{streaming-perception}, which provides the delivery-time annotations required for sAP, with the eight Argoverse-HD categories mapped to COCO categories as in Li~\etal~\cite{streaming-perception}.

To study contention among independent camera streams, we concurrently replay $N \in \{2,4,8\}$ distinct Argoverse-HD logs. This emulates independent cameras rather than a synchronized rig. For composition experiments, we sort logs by their fraction of large objects into three equal-sized groups (small-, medium-, and large-object-rich). For each multi-stream comparison, placements are evaluated using the same number of streams $N$ and the same stream composition. Worst-stream sAP is computed for each placement.

\paragraph{\textbf{Co-tenant workloads.}}
We distinguish foreground detection streams from GPU-pinned co-tenant workloads at three levels: \Lcnn{} (ResNet50~\cite{resnet}, batch~1, $224\times224$), \Llm{} (ResNet50 + TinyLlama-1.1B~\cite{tinyllama}, sequence length 32, prefill only), and \Lvlm{} (ResNet50 + Qwen2-VL-2B~\cite{qwen-vl}, pixel values $30{,}704\times1{,}176$). \Llm{} is a controlled probe rather than a representative workload. Its language model also stresses the host CPU, which helps distinguish GPU-localized contention from contention that affects the host-side stages of the NPU path. Across these three GPU co-tenant levels, inference runs at batch size 1, back-to-back on pre-generated inputs without rate limiting. All three levels execute on CUDA. We also evaluate NPU-side contention using additional ResNet50 workloads (Sec.~\ref{sec:results:reversal}).

For contention sweeps, we vary the number $k\in\{0,1,2,3,4,6,8\}$ of concurrent ResNet50 instances (additionally $k\in\{10,12,16\}$ at $N{=}2$), producing a controlled range of GPU-side contention without changing the foreground detector or placement. In the $N{=}4$ ResNet50 sweep, the four foreground streams run batch-1 YOLO11s. ResNet50 and TinyLlama run on ONNX Runtime with the CUDA provider and no explicit provider options. Qwen2-VL runs on PyTorch. These co-tenants run as threads of the foreground process without explicit CUDA streams and start $0.3$~s before the foreground streams without a readiness barrier. A ResNet50 session is preloaded in all sweep conditions, including $k{=}0$, where it performs no inference.

\paragraph{\textbf{Metrics.}}
Our primary metric is streaming AP (sAP)~\cite{streaming-perception}, which evaluates the latest available detections against the ground truth at each evaluation timestamp. Following common practice in streaming perception~\cite{streaming-perception,streamyolo}, we report sAP values in percent. We report overall and size-specific sAP. Object-size bins are defined by ground-truth box area $A$ in pixels squared, measured in original-image coordinates ($1200\times1920$): small ($A<32^2$), medium ($32^2\le A<96^2$), and large ($A\ge96^2$). For multi-stream runs, we report the mean and minimum sAP across streams, with the latter denoted as worst-stream sAP.
We also report the foreground deadline-miss rate, defined as the fraction of processed inferences whose end-to-end processing time exceeds the $33.3$~ms frame budget at 30 FPS. End-to-end processing time is measured from just before reading the image file to the completion of class filtering. For multi-stream runs, the deadline-miss rate of a path is the mean over its streams of the per-stream rate. We also record GPU utilization as a diagnostic counter.

\paragraph{\textbf{Replay protocol and repetitions.}}
Each run replays the full log on a virtual timeline (Fig.~\ref{fig:overview}b). After each inference, the stream clock advances by the measured processing time, and the latest available frame is selected next. Intervening frames are skipped without queuing. If processing finishes before the next capture, the clock advances to that capture time without sleeping. Following Li~\etal~\cite{streaming-perception}, each ground-truth frame on the 30~FPS timeline is evaluated using the latest result delivered at or before its timestamp, without delay clipping. We exclude the first $30$ processed inferences from the deadline-miss rate and the first $30$ ground-truth frames from sAP. Single-camera results are means over $24$ logs with three replays per log. Multi-stream results are repeated-run means. In the $N{=}4$ ResNet50 sweep shown in Fig.~\ref{fig:sweeps}, repetitions per placement are three at $k{=}0$, ten at $k{=}1$, seven at $k\in\{2,3\}$, and five at $k\in\{4,6,8\}$.

\paragraph{\textbf{Placement configurations and Oracle.}}
We evaluate two homogeneous placement configurations, \textbf{All-GPU} and \textbf{All-NPU}, which assign all foreground streams to the GPU or NPU pipeline, respectively. We also report a diagnostic \textbf{Oracle}, a reference selected from measured placement results rather than an online scheduler. For the ResNet50 sweep, we evaluate $N{+}1$ candidate placements. Starting from All-GPU, successive candidates assign one additional stream to the NPU in a fixed order, ending with All-NPU. At $N{=}4$, the two large-object-rich streams occupy the first two positions in this fixed order. The Oracle selects the candidate with the highest worst-stream sAP averaged over repeated runs, and its reported mean sAP is the mean sAP of that same candidate.

For \Llm{} and \Lvlm{} at $N{=}4$, we evaluate all $2^N{=}16$ placements in each repetition. Within each repetition, the Oracle selects the placement with the highest worst-stream sAP, and we then average the worst-stream and mean sAP of the selected placements over repetitions. The selected placement may differ across repetitions.

\paragraph{\textbf{Placement-selection policies.}}
Beyond the placement configurations above, we evaluate four selection policies. \emph{Isolated-accuracy selection} (P1) chooses the pipeline with higher isolated single-stream sAP. \emph{Latency-profiling selection} (P2) chooses the lowest-latency pipeline that meets the frame budget~\cite{tod}. \emph{Utilization-threshold selection} (P3) selects All-NPU when mean GPU utilization exceeds $\theta{=}80\%$ and All-GPU otherwise. \emph{Deadline-miss-aware selection} (P4) selects All-NPU when the foreground GPU deadline-miss rate exceeds $48\%$ and All-GPU otherwise. This threshold lies between two adjacent measured points ($45.0\%$ and $51.5\%$) where the ordering of All-GPU and All-NPU by worst-stream sAP reverses. We chose the threshold after examining these results and did not validate it on independent data, so P4 serves as a diagnostic policy. All policies are evaluated offline by applying each rule to recorded measurements and reporting the measured performance of the selected placement.

\begin{table}[tbp]
\centering
\caption{Single-camera evaluation with negligible deadline misses. Latency reports mean end-to-end per-frame processing time at four post-processing threads. Values are $24$-log means over three replays. Relative gaps in parentheses are computed as $(\mathrm{NPU}-\mathrm{GPU})/\mathrm{GPU}$ from the unrounded $24$-log means.}
\label{tab:single-stream}
\footnotesize
\setlength{\tabcolsep}{5pt}
\begin{tabular}{lccc}
\toprule
Metric & GPU & NPU & NPU$-$GPU \\
\midrule
end-to-end mean (ms) & 8.3 & 10.0 & --- \\
deadline misses (\%) & $\sim\!0$ & $\sim\!0$ & --- \\
sAP$_{\text{overall}}$ & 19.6 & 18.6 & $-1.0$ ($-5.0\%$) \\
sAP$_{\text{small}}$  & 1.6 & 0.8 & $-0.8$ ($-48.4\%$) \\
sAP$_{\text{medium}}$ & 18.4 & 14.8 & $-3.6$ ($-19.7\%$) \\
sAP$_{\text{large}}$  & 47.7 & 47.7 & $+0.05$ ($+0.1\%$) \\
\bottomrule
\end{tabular}
\end{table}

\subsection{Size-Dependent Accuracy Differences and sAP Losses}
\label{sec:results:single}

Table~\ref{tab:single-stream} summarizes isolated single-camera performance. Although deadline misses are near zero on both pipelines, sAP still reflects processing delays even without co-tenants. The NPU pipeline has lower sAP than the GPU pipeline by $0.8$ points ($48.4\%$) on small objects and $3.6$ points ($19.7\%$) on medium objects. Supplementary Fig.~S1 shows representative pedestrians and traffic lights missed by the NPU pipeline. For large objects, the measured sAP values are nearly identical, with the NPU pipeline higher by $0.05$ points ($0.1\%$).

Overall sAP is slightly higher on the GPU pipeline ($19.6$ vs.\ $18.6$), so isolated-accuracy profiling favors All-GPU. The GPU advantage is concentrated on small and medium objects.

\paragraph{\textbf{The same-NPU latency control shows the largest absolute sAP loss for large objects.}}
To examine how processing latency affects sAP across object sizes, we replay all logs on the same NPU pipeline with 4 and 24 host-side post-processing threads. The 24-thread setting increases end-to-end latency and produces about $55\%$ deadline misses. Detections were bit-identical for the $20$ checked input frames processed under both settings. However, the higher latency changes both which frames are processed and when results are delivered under the replay protocol. Table~\ref{tab:decomp} compares the resulting latency-control losses with the isolated GPU--NPU differences. Increasing the thread count from 4 to 24 reduces sAP by $0.1$, $3.0$, and $9.8$ points for small, medium, and large objects, respectively. These losses are consistent with latency-induced staleness. The largest absolute loss occurs for large objects, whereas the isolated GPU advantage is concentrated on small and medium objects.

\begin{table}[tb]
\centering
\caption{Size-specific sAP differences for single-stream YOLO11s. Rows compare the isolated NPU and GPU pipelines and the same NPU pipeline with 24 versus 4 post-processing threads. Differences are in sAP points. Parenthesized percentages divide these differences by the isolated GPU sAP, using $24$-log means.}

\label{tab:decomp}
\footnotesize

\setlength{\tabcolsep}{3pt}
\begin{tabular}{lccc}
\toprule
Component & Small & Medium & Large \\
\midrule
Isolated gap (NPU$-$GPU)
& $-0.8$ ($-48.4\%$)
& $-3.6$ ($-19.7\%$)
& $+0.05$ ($+0.1\%$) \\

Latency control ($24{-}4$ thr.)
& $-0.1$ ($-8.0\%$)
& $-3.0$ ($-16.1\%$)
& $-9.8$ ($-20.5\%$) \\
\bottomrule
\end{tabular}
\end{table}

\begin{figure}[tb]
\centering
\includegraphics[width=0.5\linewidth]{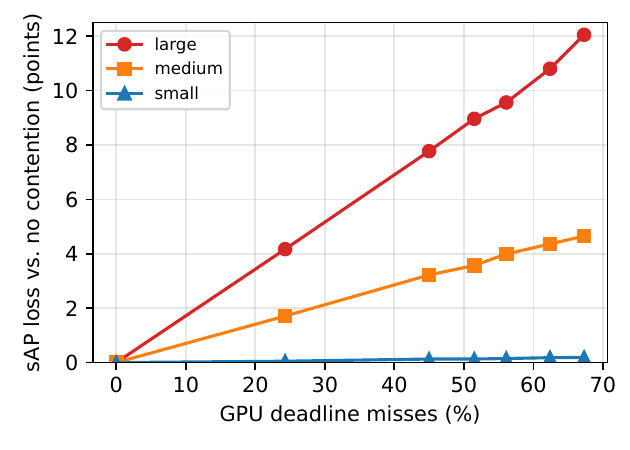}
\caption{Per-size All-GPU sAP loss under ResNet50 contention at $N{=}4$, relative to $k{=}0$. Losses are stream-averaged absolute sAP differences (points). Large-object loss rises sharply with GPU deadline misses, while small-object loss remains nearly flat.}
\label{fig:persize-sweep}
\end{figure}

\paragraph{\textbf{GPU co-tenant contention produces the same staleness pattern.}}
Figure~\ref{fig:persize-sweep} shows the same pattern on the GPU path under ResNet50 co-tenant contention. Across the sweep, large-object loss grows from $4.2$ to $12.0$ sAP points as the GPU deadline-miss rate increases from about $24\%$ to $67\%$, while small-object loss remains near zero. All-NPU runs measured under the same contention levels remain stable, with per-size sAP changing by at most $0.2$ sAP points relative to $k{=}0$, consistent with the NPU avoiding direct GPU co-tenant interference. 

\paragraph{\textbf{Detection-score differences and latency-associated sAP losses show contrasting size dependence.}}
Supplementary Fig.~S2 compares detection-score distributions for objects matched by both pipelines. The NPU-minus-GPU score differences are approximately $-0.088$, $-0.086$, and $-0.043$ for small, medium, and large objects, respectively. Among small objects matched by the GPU pipeline, $51\%$ have no NPU-pipeline detection near the object. Using only ground-truth annotations, we compare each object box with the box of the same object at later timestamps (Supplementary Fig.~S3). Small objects lose overlap fastest. The estimate combines temporal box overlap with the isolated sAP baseline for each size bin. The resulting absolute sAP loss is largest for large objects. This ordering agrees with the measured losses in the same-NPU latency control (Table~\ref{tab:decomp}) and GPU contention experiments (Fig.~\ref{fig:persize-sweep}).

\subsection{Placement Reversal Under Deployment-Time Contention}
\label{sec:results:reversal}

\paragraph{\textbf{The ranking observed in isolation can reverse under contention.}}
Table~\ref{tab:main} compares worst-stream sAP across deployment configurations. Although isolated profiling favors the GPU pipeline, the deployment-time ranking changes when contention affects the GPU path, and All-NPU becomes preferable despite its lower isolated accuracy. The single-stream analysis in Sec.~\ref{sec:problem} describes how the additional GPU-side sAP loss relative to the NPU path can outweigh the isolated GPU advantage. The multi-stream observations below are consistent with this explanation.

The clearest reversal occurs at the \Lvlm{} endpoint in Table~\ref{tab:main}, where All-GPU falls to a worst-stream sAP of $1.6$ while All-NPU maintains $8.3$. All-NPU achieves $5.2\times$ the worst-stream sAP of All-GPU, and this advantage is stable across repeated runs. In one run, the Oracle selects a mixed placement with slightly higher worst-stream sAP but lower mean sAP than All-NPU. Consequently, the reported Oracle mean sAP is lower than that of All-NPU, even though their worst-stream sAP values are both displayed as $8.3$ after rounding.

\begin{figure} [tb]
\centering
\begin{subfigure}[t]{0.49\textwidth}
\centering
\begin{tikzpicture}
\begin{axis}[
  width=\linewidth, height=0.70\linewidth,
  xlabel={GPU deadline misses (\%)}, ylabel={worst-stream sAP (\%)},
  xmin=0, xmax=84, ymin=0.0, ymax=0.122,
  xtick={0,20,40,60,80}, ytick={0.06,0.08,0.10,0.12}, yticklabels={6,8,10,12},
  legend cell align=left, legend columns=2,
  legend style={at={(0.005,0.02)}, anchor=south west, column sep=1pt, font=\fontsize{6}{7}\selectfont, draw=none, line width=0.25pt,
    inner sep=0.5pt, row sep=-2.5pt,  fill=none},
  legend image post style={scale=0.55},
  legend image code/.code={\draw[#1,mark repeat=2,mark phase=2] plot coordinates {(0mm,0mm)(1.5mm,0mm)(3mm,0mm)};},
  grid=both, major grid style={gray!18}, tick align=outside,
  label style={font=\scriptsize}, tick label style={font=\tiny},
  every axis plot/.append style={very thick, mark size=1.3pt},
]
\fill[blue!8] (axis cs:0,0.0802) rectangle (axis cs:84,0.0836);
\addplot[red,    mark=*]         coordinates {(0.0,0.1149)(3.2,0.1132)(9.0,0.1114)(12.2,0.1067)(23.0,0.1011)(34.9,0.0944)(41.3,0.0877)(52.1,0.0797)(59.3,0.0773)};
\addlegendentry{All-GPU $N{=}2$}
\addplot[blue, dashed, mark=none] coordinates {(0.0,0.0836)(24.3,0.0830)(45.0,0.0832)(51.5,0.0833)(56.1,0.0833)(62.4,0.0834)(67.3,0.0836)};
\addlegendentry{All-NPU}
\addplot[orange, mark=square*]   coordinates {(0.0,0.1149)(24.3,0.0979)(45.0,0.0868)(51.5,0.0819)(56.1,0.0788)(62.4,0.0767)(67.3,0.0713)};
\addlegendentry{All-GPU $N{=}4$}
\addplot[brown,  mark=triangle*] coordinates {(58.9,0.0714)(67.0,0.0567)(68.8,0.0590)(70.6,0.0584)(72.8,0.0584)(76.5,0.0564)(79.6,0.0541)};
\addlegendentry{All-GPU $N{=}8$}
\end{axis}
\end{tikzpicture}
\caption{Homogeneous placements for $N{=}2,4,8$. The dashed All-NPU line is for $N{=}4$. The shaded band spans the measured All-NPU values across $N{=}2,4,8$ ($k\in\{0,8\}$ only for $N{=}2,8$).}
\label{fig:sweep}
\end{subfigure}
\hfill
\begin{subfigure}[t]{0.49\textwidth}
\centering
\begin{tikzpicture}
\begin{axis}[
  width=\linewidth, height=0.70\linewidth,
  xlabel={GPU deadline misses (\%)}, ylabel={worst-stream sAP (\%)},
  xmin=0, xmax=72, ymin=0.02, ymax=0.122,
  xtick={0,20,40,60}, ytick={0.06,0.08,0.10,0.12}, yticklabels={6,8,10,12},
  legend pos=south west, legend cell align=left, legend columns=2,
  legend style={font=\fontsize{6}{7}\selectfont, draw=none, line width=0.25pt, inner sep=0.5pt, row sep=-2.5pt,  fill=none},
  legend image post style={scale=0.55},
  legend image code/.code={\draw[#1,mark repeat=2,mark phase=2] plot coordinates {(0mm,0mm)(1.5mm,0mm)(3mm,0mm)};},
  grid=both, major grid style={gray!18}, tick align=outside,
  label style={font=\scriptsize}, tick label style={font=\tiny},
  every axis plot/.append style={very thick, mark size=1.5pt},
]
\addplot[orange, mark=square*] coordinates {(0.0,0.1149)(24.3,0.0979)(45.0,0.0868)(51.5,0.0819)(56.1,0.0788)(62.4,0.0767)(67.3,0.0713)};
\addlegendentry{All-GPU}
\addplot[blue, mark=*] coordinates {(0.0,0.0836)(24.3,0.0830)(45.0,0.0832)(51.5,0.0833)(56.1,0.0833)(62.4,0.0834)(67.3,0.0836)};
\addlegendentry{All-NPU}
\addplot[teal, mark=diamond*, densely dashed] coordinates {(0.0,0.1149)(24.3,0.1026)(45.0,0.0868)(51.5,0.0836)(56.1,0.0836)(62.4,0.0836)(67.3,0.0836)};
\addlegendentry{Oracle}
\node[circle,draw=teal,thick,inner sep=2.2pt] at (axis cs:24.3,0.1026){};
\node[anchor=south west,teal,font=\tiny] at (axis cs:26,0.1038){Oracle $>$ both};
\end{axis}
\end{tikzpicture}
\caption{All-GPU, All-NPU, and diagnostic Oracle under mixed placements, $N{=}4$.}
\label{fig:split}
\end{subfigure}
\caption{Worst-stream sAP versus GPU deadline misses under GPU-localized ResNet50 contention. All-GPU degrades as GPU deadline misses increase, while All-NPU remains stable. The ranking of All-GPU and All-NPU reverses between the adjacent measured points at $45.0\%$ and $51.5\%$ GPU deadline misses ($N{=}4$), before full GPU saturation.}

\label{fig:sweeps}
\end{figure}
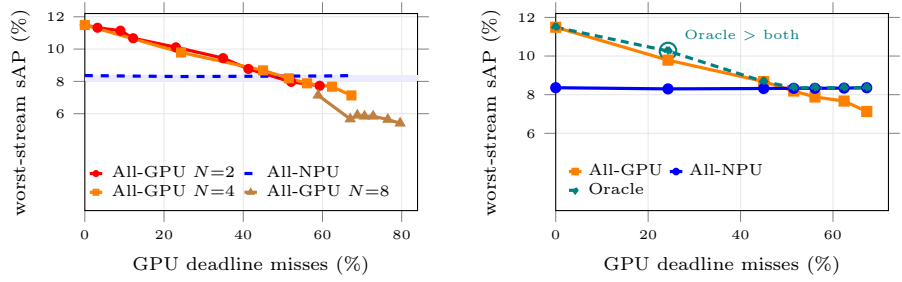

Figure~\ref{fig:sweeps} shows that this reversal appears before the GPU-saturating \Lvlm{} endpoint. In the ResNet50 sweep, All-GPU worst-stream sAP decreases as GPU deadline misses increase, while All-NPU remains nearly constant. The worst-stream sAP ranking of All-GPU and All-NPU reverses between the adjacent measured GPU deadline-miss rates of $45.0\%$ and $51.5\%$, so complete GPU saturation is not required. To test this crossover, we repeat the measurements at $k\in\{1,2,3\}$ using ten runs per placement. All-GPU and All-NPU runs are paired by repetition index and log set. At $k{=}2$ (mean GPU deadline-miss rate $45.2\%$), the mean paired difference in worst-stream sAP (All-GPU minus All-NPU) is $+0.182$ points (95\% paired $t$ CI $[+0.009,+0.354]$). At $k{=}3$ ($53.8\%$), it is $-0.453$ points (95\% CI $[-0.570,-0.337]$). The two intervals lie on opposite sides of zero, supporting a reversal between these settings. Because we vary the discrete co-tenant count $k$ rather than the deadline-miss rate, $45.2$--$53.8\%$ brackets the reversal but is not a confidence interval for its location.

\begin{table}[tbp]
\centering
\caption{Worst-stream and mean sAP under GPU co-tenancy at $N{=}4$. GPU/NPU DM denotes the GPU deadline-miss rate under All-GPU and the NPU deadline-miss rate under All-NPU (\%).}
\label{tab:main}
\footnotesize
\setlength{\tabcolsep}{5pt}
\begin{tabular}{lcc|cc|cc|c}
\toprule
 & \multicolumn{2}{c|}{All-GPU} & \multicolumn{2}{c|}{All-NPU} & \multicolumn{2}{c|}{Oracle} & GPU/NPU \\
Co-tenant & worst & mean & worst & mean & worst & mean & DM (\%) \\
\midrule
\Lcnn{} (ResNet50)   & $9.8$ & $13.7$ & $8.3$ & $12.6$ & $10.3$ & $13.7$ & $24 / 1$ \\
\Llm{} ($+$LLM)   & $6.2$ & $10.0$ & $4.2$ & $8.8$ & $6.3$ & $9.9$ & $83 / 76$ \\
\Lvlm{} ($+$VLM)  & $1.6$ & $4.7$ & $8.3$ & $12.6$ & $8.3$ & $12.0$ & $100 / 1$ \\
\bottomrule
\end{tabular}
\end{table}

\paragraph{\textbf{The reversal depends on the bottlenecked path.}}
Table~\ref{tab:main} shows that, in the tested conditions, placement reversal depends on which execution path becomes stale, not on contention magnitude alone. Under the light CNN co-tenant, GPU placements remain competitive with a mean GPU deadline-miss rate of $24\%$. Under the \Llm{} co-tenant, GPU and NPU deadline-miss rates reach $83\%$ and $76\%$, respectively. CPU utilization rises from $17\%$ to $93\%$. The NPU pipeline also relies on the host for image decoding, letterbox pre-processing, and post-processing, so shared-host contention can delay this path. All-NPU worst-stream sAP falls from $8.3$ under \Lcnn{} to $4.2$ under \Llm{}. All-GPU retains higher worst-stream sAP ($6.2$ vs.\ $4.2$), and the Oracle offers only a small improvement over All-GPU. Under \Lvlm{}, by contrast, the GPU deadline-miss rate reaches $100\%$ while the NPU rate remains near $1\%$. All-NPU then outperforms All-GPU.

\paragraph{\textbf{NPU-side contention did not reverse the ranking under the tested load.}}
We run four foreground streams and add NPU-side contention using eight host threads that issue ResNet50 requests to the NPU. Each placement is evaluated over ten repetitions. Under this load, foreground NPU end-to-end latency increases from $17.2$ to $26.9$~ms. The mean paired latency increase is $9.7$~ms (95\% CI $[9.4,9.9]$). Despite this increase, the mean per-stream NPU deadline-miss rate is only $1.5\%$. Without co-tenants, All-GPU leads All-NPU by $3.14$ worst-stream sAP points. Under NPU-side load, All-GPU still leads by $3.19$ points. Under GPU-side ResNet50 load ($k{=}8$), by contrast, the mean per-stream GPU deadline-miss rate reaches $71.7\%$, while NPU latency remains close to its no-co-tenant value ($17.3$ vs.\ $17.2$~ms). In this condition, the ranking reverses, and All-NPU leads All-GPU by $1.45$ worst-stream sAP points. The GPU-side and NPU-side loads are not matched in intensity, so these results do not establish how the placements would rank under equally severe contention on the two paths.

\begin{table}[tbp]
\centering
\caption{Worst-stream sAP of placement-selection policies at $N{=}4$. Parentheses report Oracle minus policy sAP in points. Gaps are computed from unrounded values. Row percentages denote All-GPU deadline-miss rates. Superscripts G and N indicate All-GPU and All-NPU. Policies and Oracle are defined in Sec.~\ref{sec:setup}.}
\label{tab:policy-comparison}
\footnotesize
\setlength{\tabcolsep}{7pt}
\begin{tabular}{lcccc}
\toprule
Contention point & P1/P2 & P3 & P4 & Oracle \\
\cmidrule(lr){2-4} \cmidrule(lr){5-5}
 & \multicolumn{3}{c}{sAP (gap to Oracle)} & sAP \\
\midrule
\Lcnn{} (24\%) & 9.8 (+0.5) & 9.8 (+0.5)$^{\mathrm{G}}$ & 9.8 (+0.5)$^{\mathrm{G}}$ & 10.3 \\
ResNet (52\%) & 8.2 (+0.2) & 8.2 (+0.2)$^{\mathrm{G}}$ & 8.3 (+0.0)$^{\mathrm{N}}$ & 8.4 \\
ResNet (67\%) & 7.1 (+1.2) & 7.1 (+1.2)$^{\mathrm{G}}$ & 8.4 (+0.0)$^{\mathrm{N}}$ & 8.4 \\
\Llm{} (83\%) & 6.2 (+0.1) & 6.2 (+0.1)$^{\mathrm{G}\dagger}$ & 4.2 (+2.0)$^{\mathrm{N}}$ & 6.3 \\
\Lvlm{} (100\%) & 1.6 (+6.8) & 8.3 (+0.0)$^{\mathrm{N}}$ & 8.3 (+0.0)$^{\mathrm{N}}$ & 8.3 \\
\bottomrule
\multicolumn{5}{p{0.93\linewidth}}{\scriptsize P1 uses isolated sAP and P2 isolated latency. P3 applies an $80\%$ threshold to GPU utilization. P4 applies a $48\%$ threshold to the GPU deadline-miss rate. P1 and P2 select the same placement in all rows. $^\dagger$Under \Llm{}, GPU utilization stays below $80\%$, so P3 selects All-GPU.} \\

\end{tabular}
\end{table}
\paragraph{\textbf{Limitations of placement-selection criteria.}}
Table~\ref{tab:policy-comparison} reports the placement each selection criterion chooses at representative contention levels, with its worst-stream sAP gap to the Oracle. Isolated-accuracy (P1) and latency-profiling (P2) selection always choose All-GPU, since the GPU wins both isolated sAP ($19.6$ vs.\ $18.6$) and isolated latency ($8.3$~ms vs.\ $10.0$~ms). Their gap to the Oracle reaches $1.2$ sAP points at the highest ResNet50 contention level and $6.8$ points under \Lvlm{}.

Utilization-threshold selection (P3) fails differently. Mean GPU utilization stays near $68\%$ while deadline misses climb from $24\%$ to $67\%$, so the $80\%$ threshold is never crossed during the sweep (nor are $70\%$ or $90\%$), and P3 triggers only when \Lvlm{} fully saturates the GPU. Utilization is thus a lagging signal, whereas the reversal appears mid-sweep. Deadline-miss-aware selection (P4) achieves near-Oracle sAP at the higher ResNet50 contention levels but fails under \Llm{}, where host-CPU contention also stalls the NPU path. Using only GPU deadline misses, P4 selects All-NPU and incurs a $+2.0$ sAP gap, showing that deadline misses on both paths must be checked.

\paragraph{\textbf{QoS mitigates contention but does not remove the reversal.}}
\label{sec:results:qos}

A natural objection is that the reversal merely reflects unmanaged scheduling. We therefore assign all foreground inference to a maximum-priority CUDA stream and repeat the $N{=}4$ sweep. QoS improves foreground latency and worst-stream sAP but does not restore uncontended behavior. Under the GPU-saturating VLM co-tenant, All-NPU (worst-stream sAP $8.3$) still outperforms protected All-GPU ($3.3$) by $2.5\times$. QoS shifts the crossover to heavier contention but does not eliminate the reversal in the tested VLM condition.

\subsection{Mean sAP Understates the Worst Camera}
\label{sec:results:worst}
Table~\ref{tab:main} shows a substantial gap between mean and worst-stream sAP under contention. Under the VLM co-tenant, All-GPU achieves a mean sAP of $4.7$ but a worst-stream sAP of only $1.6$ (mean-to-worst ratio $2.9$), whereas All-NPU shows a smaller mean-to-worst ratio of $1.5$ ($12.6$ vs.\ $8.3$).

Supplementary Table~S1 shows the same masking in a single replay at a high-contention ResNet50 sweep point with $N{=}4$. All-GPU has a mean sAP of $10.7$ and a median of $8.7$, both above the worst-stream sAP of $6.6$. The 10th-percentile sAP of $7.2$ is closer to the worst-stream value. Excluding the worst stream from the average raises the mean from $10.7$ to $12.1$. Central-tendency metrics therefore understate failures concentrated in a single stream, which matters because one severely degraded camera can compromise the perception system. Lower-tail metrics such as lower percentiles, CVaR, and min-$k$ averages could also expose this degradation. We use worst-stream sAP because it directly reports the performance of the weakest camera and because percentile and CVaR estimates become coarse at small $N$.

Notably, the worst-stream score is not determined only by stream content. The same stream is usually the worst across placements. Under GPU-localized contention, its sAP drops sharply on the GPU but remains nearly unchanged under All-NPU. Worst-stream sAP thus captures contention-induced degradation, not just differences among streams.

\section{Discussion}
\label{sec:discussion}

\subsection{Implications for Streaming-Perception Evaluation}
\label{sec:discussion:evaluation}

\paragraph{\textbf{A deployment-faithful reporting protocol.}}
We distill our findings into three reporting recommendations. First, report sAP under co-tenancy localized to the shared accelerator as well as in isolation. Second, report worst-stream sAP alongside mean sAP. Third, report foreground deadline-miss rates on both the GPU and NPU paths across the contention sweep and identify the measured range in which the ranking reverses. Table~\ref{tab:policy-comparison} shows that relying on GPU deadline misses alone mis-selects the placement under the \Llm{} co-tenant. Reporting only isolated sAP, mean sAP, and aggregate GPU utilization at a single operating point can hide both the reversal and its severity.

\subsection{Limitations}
\label{sec:discussion:limitations}
 
\noindent\textbf{Detector and toolchain coverage.}
Table~\ref{tab:detector-generality} shows that the same reversal appears on YOLOv8n, YOLOv8s, YOLOv10s, and YOLO12s. The large-object isolated gap and crossover bracket vary by detector. We have not tested larger models or non-YOLO detectors. Supplementary Table~S2 reports GPU--NPU comparisons for COCO-pretrained and fine-tuned YOLO11s under a local NPU compile recipe. Both models show a statistically significant small-object sAP deficit of the NPU pipeline (relative gaps of $-43.1\%$ and $-34.9\%$, respectively, both $p<0.001$) and no significant large-object gap. These results are not directly comparable to Table~\ref{tab:single-stream} and do not test the effect of fine-tuning. The sAP-loss estimate matches the observed ordering of per-size losses but not their magnitudes, and the confidence analysis covers only objects that the GPU pipeline localizes.

\noindent\textbf{When object size reflects image motion.}
Our size-based analysis relies on a common pattern in ego-motion driving scenes such as Argoverse-HD, where larger objects tend to be closer and move faster in the image. This link does not always hold. In scenes where size and image motion are weakly linked, such as fixed surveillance cameras, size may not predict staleness well, and grouping objects by image-plane velocity would be more direct.

\begin{table}[tb]
\centering
\caption{Placement reversal across detectors. Isolated columns report pipeline preference and relative sAP differences $(\mathrm{NPU}-\mathrm{GPU})/\mathrm{GPU}$ computed from $24$-log means. ``Within 5\%'' denotes numerical proximity, not statistical equivalence. Under VLM contention, ``VLM cont.'' identifies the preferred pipeline, and ``worst NPU/GPU'' reports worst-stream sAP for All-NPU and All-GPU, respectively. GPU-DM brackets give adjacent measured GPU deadline-miss rates between which the All-GPU versus All-NPU worst-stream sAP ordering reverses. They are not interpolated estimates or confidence intervals.}
\label{tab:detector-generality}
\scriptsize
\setlength{\tabcolsep}{3pt}
\begin{tabular}{lccccc}
\toprule
Detector & isolated small & isolated large & VLM cont. & worst NPU/GPU & GPU-DM bracket \\
\midrule
YOLO12s  & GPU ($-36\%$) & within 5\% ($-1.2\%$) & NPU & $8.0/1.1$ & $0.4$--$61.5\%$ \\
YOLO11s & GPU ($-48\%$) & within 5\% ($+0.1\%$) & NPU & $8.3/1.6$ & $45.0$--$51.5\%$ \\
YOLOv10s & GPU ($-45\%$) & within 5\% ($-4.6\%$) & NPU & $8.2/1.6$ & $0.0$--$29.7\%$ \\
YOLOv8s  & GPU ($-44\%$) & within 5\% ($-0.3\%$) & NPU & $8.0/1.5$ & $10.8$--$32.4\%$ \\
YOLOv8n  & GPU ($-36\%$) & GPU ($-8.1\%$) & NPU & $6.0/0.9$ & $34.3$--$46.9\%$ \\
\bottomrule
\end{tabular}
\end{table}
 
\noindent\textbf{Measurement stability.}
Unless otherwise stated, sAP, latency, and deadline-miss measurements are averaged over at least three repeats (ten for the crossover re-measurement and the NPU-side contention experiment). Under the \Lvlm{} co-tenant, the standard deviation of worst-stream sAP is below $0.1$ points for both All-GPU and All-NPU across three repeats. The light \Lcnn{} co-tenant at $N{=}4$ operates near the $33.3$~ms deadline. Over ten repeats, its GPU deadline-miss rate was $24.3 \pm 2.2\%$ (mean $\pm$ SD). The first mixed candidate of Sec.~\ref{sec:setup} assigns one large-object-rich stream to the NPU and the remaining streams to the GPU. This candidate had the highest worst-stream sAP in nine of ten runs and achieved a worst-stream sAP of $10.3 \pm 0.3$, compared with $9.8 \pm 0.4$ for All-GPU. This variability therefore does not alter the observed pattern.

\noindent\textbf{Pipeline confound.}
Our comparison does not separate device, precision, and software-stack effects. The claims therefore apply to the evaluated end-to-end pipelines, not the devices. In a precision sensitivity check using GPU TensorRT INT8 on the same $640\times640$ canvas, the NPU pipeline achieves higher isolated sAP on all four evaluated logs. The reported isolated ordering therefore does not hold in this check. Supplementary Table~S3 and the accompanying text report the per-log results and calibration details. This check matches nominal precision and input canvas but does not match calibration data or the full software stack. NPU calibration data are unavailable. The results apply only to the tested post-training quantization configuration.
 
\noindent\textbf{Platform and workload scope.}
We evaluate a single NPU vendor and card, and emulate multi-camera workloads by replaying distinct single-camera logs. The degree of NPU isolation from GPU contention depends on the platform. SoC-integrated NPUs may share memory bandwidth with the GPU. Because idle time is skipped rather than slept, foreground streams issue work back-to-back in wall-clock time at low load, so interference may differ from a wall-clock-paced 30~FPS deployment. Validation on synchronized multi-camera benchmarks with streaming labels remains future work.

\section{Conclusion}
\label{sec:conclusion}

Isolated streaming-perception evaluation can mis-rank placement on a shared GPU--NPU platform. For the two evaluated end-to-end pipelines, the GPU pipeline is preferred in isolation, but GPU-localized contention makes its detections stale and can make the NPU pipeline the better deployment-time choice before full GPU saturation. The ranking depends on which execution path becomes stale, not on contention magnitude alone. The multi-stream results are consistent with an account in which the staleness-loss difference between the two paths exceeds the measured isolated pipeline gap. The reversal reproduces across five YOLO variants and persists under maximum-priority CUDA streams in the tested VLM condition. The mechanism may extend to other platforms, but crossover values and placement decisions remain hardware dependent. These results show that isolated accuracy, isolated latency, and aggregate GPU utilization alone are insufficient for placement selection on the evaluated platform. Placement evaluation should measure streaming accuracy across deployment-time contention levels and report deadline-miss rates on both paths together with mean and worst-stream sAP.

\subsubsection*{\ackname}
This work was supported by Institute of Information \& communications Technology Planning \& Evaluation (IITP) grants funded by the Korea government (MSIT) (No.~RS-2023-00277060, Development of open edge AI SoC hardware and software platform; IITP-2026-RS-2026-25615995, AI Star Fellowship Support Program; RS-2020-II200004, Development of Previsional Intelligence based on Long-term Visual Memory Network) and by the Electronics and Telecommunications Research Institute (ETRI) grant funded by the Korean government [26CS1100, Development of Proprietary Physical AI-based Small-scale Computers and Integrated Soft Suits].

\clearpage
\raggedbottom
\section*{Supplementary Material}
\label{sec:supplementary}

This supplement collects supporting figures and tables referenced from the main paper. Figures and tables here are numbered S1, S2, \ldots. References without an ``S'' prefix (e.g., Table~2, Sec.~4.2) refer to the main paper.

\setcounter{section}{0}
\setcounter{figure}{0}
\setcounter{table}{0}
\renewcommand{\thefigure}{S\arabic{figure}}
\renewcommand{\thetable}{S\arabic{table}}
\renewcommand{\thesection}{S\arabic{section}}
\renewcommand{\theHfigure}{S\arabic{figure}}
\renewcommand{\theHtable}{S\arabic{table}}
\renewcommand{\theHsection}{S\arabic{section}}

\section{Qualitative Differences Between the Two Pipelines (Sec.~4.2)}

\begin{figure}[H]
\centering
\includegraphics[width=0.92\linewidth]{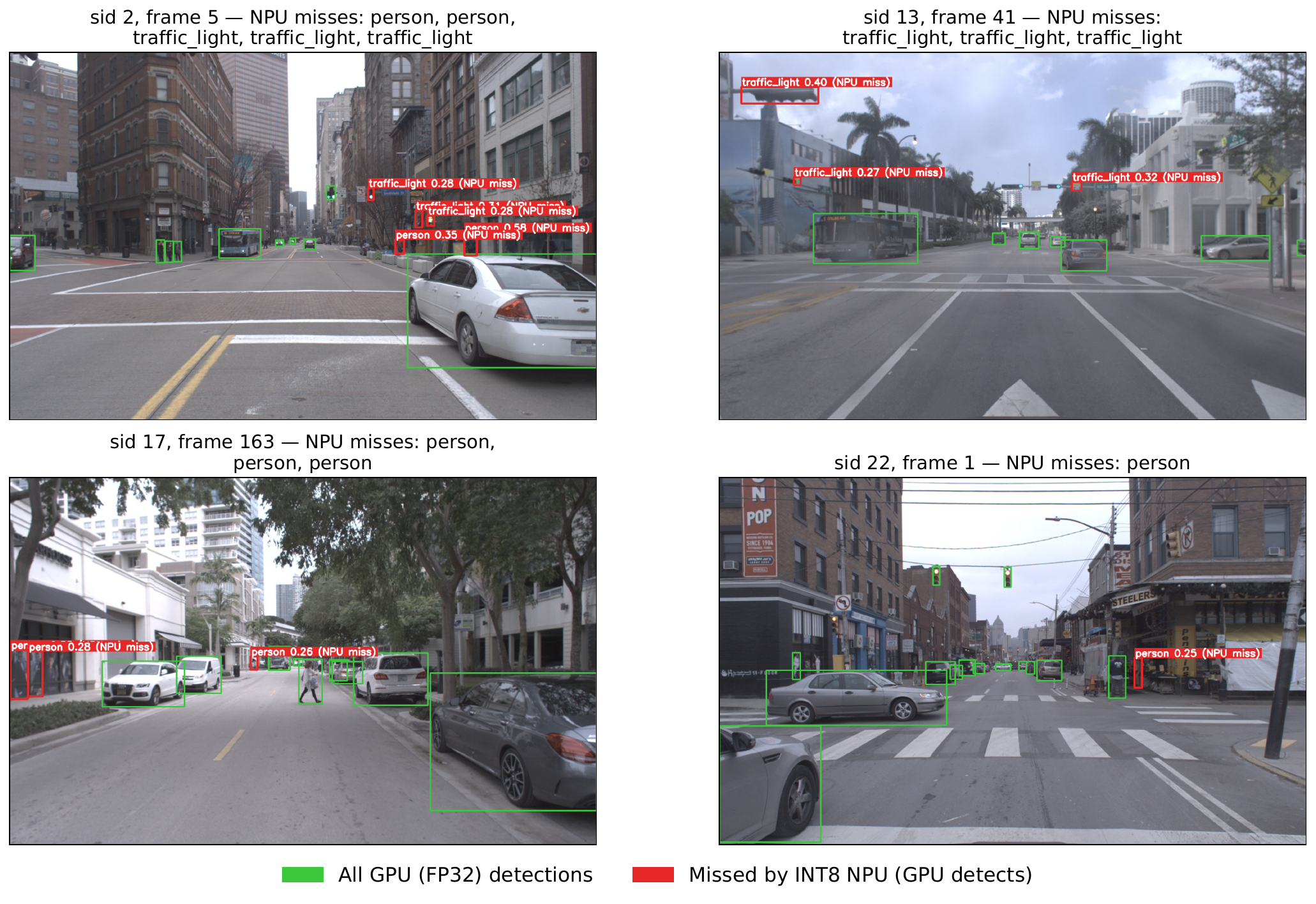}
\caption{Qualitative differences between the two evaluated pipelines (YOLO11s, Argoverse-HD). Green: GPU-pipeline detections. Red: detections missed by the NPU pipeline. Missed detections are predominantly small and distant objects.}
\label{fig:s-failures}
\end{figure}

\section{Per-Size Confidence Distributions (Sec.~4.2)}

We match detections of the two pipelines against the ground truth and compare matched-detection scores. Detections are matched to ground truth within each class by greedy assignment at IoU $\ge 0.5$, independently for each pipeline, with a detection threshold of $0.25$. The histograms include only objects matched by both pipelines, and score differences are pooled over all matched pairs across the $24$ logs. The analysis covers only objects that the GPU pipeline localizes. For $4{,}525$ of the $8{,}818$ small objects matched by the GPU pipeline, the best IoU with any same-class NPU-pipeline detection is below $0.1$. These objects account for approximately $51\%$ of the GPU-matched small objects and are excluded from the histograms.

\begin{figure}[H]
\centering
\includegraphics[width=\linewidth]{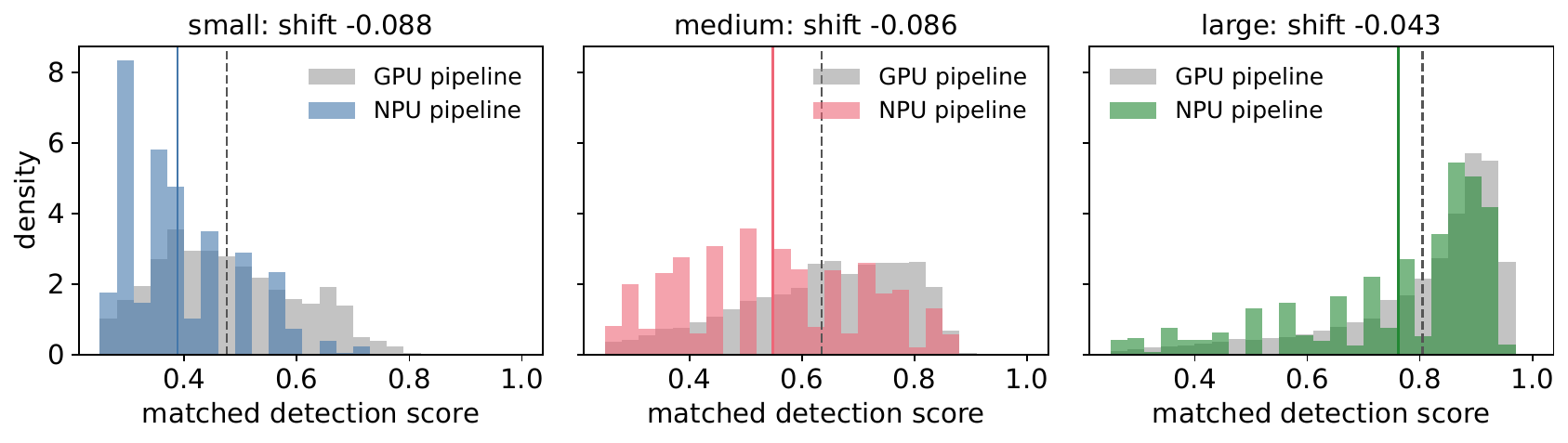}
\caption{Per-size confidence distributions for detections matched by both the GPU and NPU pipelines. Vertical lines mark the mean score for each pipeline (dashed: GPU, solid: NPU), and the panel title reports the mean NPU-minus-GPU score difference. The NPU pipeline shifts scores downward, especially for small and medium objects. Small objects with no nearby NPU-pipeline detection ($51\%$ of the GPU-matched small objects) are absent from the plotted distributions. The pattern is consistent with quantization but is not separated from other software-stack effects.}
\label{fig:s-quant-conf}
\end{figure}

\section{Temporal Box Overlap and Estimated sAP Loss (Sec.~4.2)}

\begin{figure}[H]
\centering
\includegraphics[width=\linewidth]{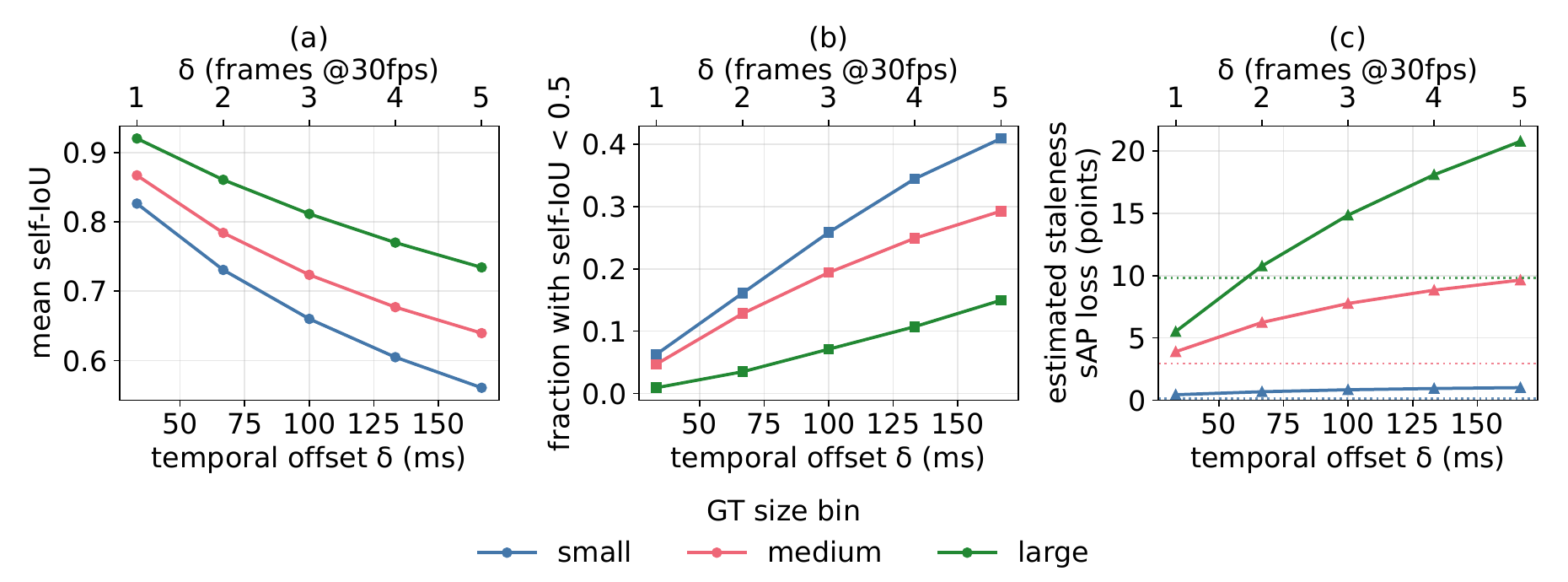}
\caption{Temporal box overlap and estimated sAP loss using Argoverse-HD ground truth. (a) Mean same-object IoU versus temporal offset $\delta$. (b) Fraction of objects with same-object IoU below $0.5$. (c) Estimated staleness sAP loss (solid) versus the measured same-NPU latency-control losses from Table~2 of the main paper (dotted). The vertical axis in panel (c) is in absolute sAP points. Box overlap decreases fastest for small objects, but the estimated sAP loss is largest for large objects. The estimated losses show the same size ordering as the measured losses but do not reproduce their magnitudes.}
\label{fig:s-iou-decay}
\end{figure}

Panel (c) of Fig.~\ref{fig:s-iou-decay} estimates sAP loss using the same ground-truth box pairs as panels (a) and (b). For each size bin and delay $\delta$, we pair every ground-truth box of the same Argoverse-HD track at frames $t$ and $t+\delta$, excluding boxes marked ignore or crowd. The size bin is set by the frame-$t$ box area using the COCO thresholds ($32^2$ and $96^2$) on native $1920\times1200$ pixels. Delays are $\delta\in\{1,\ldots,5\}$ frames at 30~FPS.

For each size bin and delay, we calculate the fraction of box pairs with IoU below each of the ten thresholds from $0.50$ to $0.95$ in steps of $0.05$. We average these fractions and multiply the result by the isolated GPU-pipeline sAP for that size bin to estimate the sAP loss. The baseline sAP values are measured with four post-processing threads and reported in Table~1 of the main paper.

\section{Metric Sensitivity (Sec.~4.4)}

\begin{table}[H]
\centering
\caption{Metric sensitivity at the ResNet50 $k{=}8$ sweep point ($N{=}4$, four post-processing threads, GPU deadline-miss rate about $80\%$, single replay). Mean and median mask the degraded camera, whereas worst-stream and lower-tail statistics expose it. Percentiles use linear interpolation. With only four per-stream values, the 10th percentile is coarse and is closer to the worst-stream value than the mean or median for both placements. Removing the worst camera raises the All-GPU system mean from $10.7$ to $12.1$.}
\label{tab:s-metric-sensitivity}
\footnotesize
\setlength{\tabcolsep}{8pt}
\begin{tabular}{lcc}
\toprule
statistic & All-GPU & All-NPU \\
\midrule
mean      & $10.7$ & $12.6$ \\
median    & $8.7$ & $11.5$ \\
$p_{10}$  & $7.2$ & $9.1$ \\
worst     & $6.6$ & $8.4$ \\
\bottomrule
\end{tabular}
\end{table}

\clearpage
\section{Fine-Tuned Detector Under a Local Compile Recipe (Sec.~5.2)}

Table~\ref{tab:s-ft} compares the GPU and NPU pipelines when the NPU model is produced by a local compile recipe rather than the vendor pipeline used in the main paper. Calibration uses $200$ images from the Argoverse-HD train split, and validation logs are never used for calibration. The same recipe is applied identically to COCO-pretrained and Argoverse-HD fine-tuned YOLO11s. Measurements are isolated single-stream runs over $24$ logs with three repeats and deadline misses near zero. For each model condition and size bin, we compare GPU and NPU sAP using a two-sided paired Wilcoxon test over per-log means of three repeats ($n{=}24$). The $p$-values in the table are uncorrected. The four significant cells remain significant under Bonferroni correction ($\alpha=0.05/6$), and the two large-object cells remain non-significant. The tests do not assess the effect of fine-tuning, and the values are not directly comparable to Table~1 of the main paper.

\begin{table}[H]
\centering
\caption{Per-size relative sAP gap $(\mathrm{NPU}-\mathrm{GPU})/\mathrm{GPU}$ under a single local compile recipe applied to COCO-pretrained and Argoverse-HD fine-tuned YOLO11s. The magnitude of the relative sAP gap is largest for small objects in both model conditions. The values are an internal paired comparison. n.s.\ denotes not statistically significant ($p \ge 0.05$).}
\label{tab:s-ft}
\footnotesize
\begin{tabular}{llll}
\toprule
weights & small & medium & large \\
\midrule
COCO-pretrained & $-43.1\%$ ($p<0.001$) & $-16.5\%$ ($p<0.001$) & $+0.6\%$ (n.s., $p=0.33$) \\
fine-tuned      & $-34.9\%$ ($p<0.001$) & $-16.7\%$ ($p<0.001$) & $-4.3\%$ (n.s., $p=0.079$) \\
\bottomrule
\end{tabular}
\end{table}

\clearpage
\section{Precision Sensitivity Check: GPU TensorRT INT8 (Sec.~5.2)}

Table~\ref{tab:s-trt} reports the per-log values of the precision sensitivity check described in the main paper. Both pipelines use the same $640\times640$ canvas without co-tenants. The GPU engine uses TensorRT INT8 with entropy-calibrated post-training quantization. Calibration uses $200$ unlabeled images drawn from validation logs other than the four evaluated here. NPU calibration data are unavailable. Values are means over nine repetitions per pipeline per log. The NPU pipeline leads in all $36$ paired repetitions across the four logs. The check matches nominal precision and input canvas but not calibration data or the full software stack. The results apply only to the tested post-training quantization configuration.

\begin{table}[H]
\centering
\caption{Single-stream overall sAP in percent per evaluated log, mean over nine repetitions. NPU$-$GPU differences are in sAP points and are computed from unrounded means. Deadline misses are $\approx0$ for both pipelines (mean end-to-end latency $14.3$~ms NPU, $6.8$~ms GPU TensorRT INT8).}
\label{tab:s-trt}
\footnotesize
\setlength{\tabcolsep}{7pt}
\begin{tabular}{lccccc}
\toprule
Pipeline & log 2 & log 22 & log 3 & log 21 & mean \\
\midrule
NPU (vendor INT8)        & $10.90$ & $19.12$ & $12.14$ & $8.35$ & $12.63$ \\
GPU (TensorRT INT8)      & $8.99$  & $10.30$ & $6.87$  & $6.37$ & $8.13$ \\
NPU$-$GPU (per-log mean) & $+1.91$ & $+8.83$ & $+5.28$ & $+1.99$ & $+4.50$ \\
\bottomrule
\end{tabular}
\end{table}

The log numbers denote the following Argoverse-HD validation sequences. The first two are small-object-rich and the last two are large-object-rich.

\begin{center}
\footnotesize
\begin{tabular}{ll}
\toprule
log & Argoverse-HD sequence \\
\midrule
2  & \texttt{1d676737-4110-3f7e-bec0-0c90f74c248f} \\
22 & \texttt{f1008c18-e76e-3c24-adcc-da9858fac145} \\
3  & \texttt{2d12da1d-5238-3870-bfbc-b281d5e8c1a1} \\
21 & \texttt{e9a96218-365b-3ecd-a800-ed2c4c306c78} \\
\bottomrule
\end{tabular}
\end{center}

\bibliographystyle{splncs04}
\bibliography{main}

\end{document}